\documentclass[conference]{IEEEtran}
\usepackage{trackchanges}
\addeditor{vg}
\addeditor{dz}
\addeditor{zahra}

\ifCLASSINFOpdf
\else

\fi

\usepackage{dingbat}

\usepackage[ruled]{algorithm2e}
\usepackage[noend]{algpseudocode}
\usepackage{stmaryrd}
\usepackage{marvosym}
\usepackage{booktabs}
\usepackage{enumitem}
 \usepackage{moresize}
\usepackage{dblfloatfix}  
\usepackage{varwidth}
\usepackage{capt-of}
\usepackage{listings}
\usepackage{amsmath}
\usepackage{framed}
\usepackage{tabu}
\usepackage{float}
\usepackage[caption = false]{subfig}
\usepackage{enumitem}
\usepackage{color}
\usepackage{graphicx}
\usepackage{cleveref}
\usepackage{fancybox}
\usepackage[export]{adjustbox}
\usepackage[table]{xcolor}
\definecolor{Gray}{gray}{0.9}
\definecolor{White}{RGB}{255,255,255}
\usepackage{bbold}
\usepackage{bbm}
\usepackage{mathbbol}
\usepackage{multirow}
\usepackage{fix-cm}
\SetKwInput{KwData}{Input}
\usepackage{siunitx,booktabs}
\algnewcommand\algorithmicforeach{\textbf{for each}}
\algdef{S}[FOR]{ForEach}[1]{\algorithmicforeach\ #1\ \algorithmicdo}
\usepackage{xcolor}
\def\headline#1{\hbox to \hsize{\hrulefill\quad\lower.3em\hbox{#1}\quad\hrulefill}}
\usepackage{setspace}
\usepackage{graphicx}
\usepackage{colortbl}
\usepackage{array}
\usepackage{pifont}
\usepackage {rotating}
\usepackage{mwe}
\usepackage{pbox}
\usepackage{enumitem}
\let\oldding\ding
\renewcommand{\ding}[2][1]{\scalebox{#1}{\oldding{#2}}}
\setlist[description]{leftmargin=\parindent,labelindent=\parindent}
\newcommand{\Csharp}{%
  {\settoheight{\dimen0}{C}C\kern-.05em \resizebox{!}{\dimen0}{\raisebox{\depth}{\#}}}}

 \usepackage{enumitem}
\usepackage{xspace}
\usepackage{xparse}
\DeclareDocumentCommand\newstep{o}{%
\item\IfNoValueTF{#1}{}{#1 \textendash\xspace}}

\newlist{steps}{enumerate}{1}
\setlist[steps]{label=\textit{Step \arabic*:},leftmargin=*}
 \usepackage[style=ieee, backend=bibtex, defernumbers=true]{biblatex}

\usepackage{fontawesome}
\definecolor{orange}{RGB}{0,32,96}
\definecolor{g}{RGB}{50,50,50}
\definecolor{brightmaroon}{rgb}{0.76, 0.13, 0.28}
\definecolor{byzantine}{rgb}{0.74, 0.2, 0.64}
\definecolor{chromeyellow}{rgb}{1.0, 0.65, 0.0}
\definecolor{applegreen}{rgb}{0.55, 0.5, 0.0}
\definecolor{cadetgrey}{rgb}{0.57, 0.64, 0.69}

\usepackage{capt-of}

\PassOptionsToPackage{gray}{xcolor}
\usepackage{tikz}

\usepackage{tcolorbox}

\makeatletter
\def\ps@IEEEtitlepagestyle{%
  \def\@oddfoot{\mycopyrightnotice}%
  \def\@evenfoot{}%
}
\def\mycopyrightnotice{%
  {\hfill \footnotesize [Preprint version] Accepted to the 26$^{\text{th}}$ IEEE International Conference on Requirements Engineering (RE'18) \hfill}
}
\makeatother

\begin{document}

\title{Dynamic Visual Analytics for Elicitation Meetings with ELICA}

\author{
    \IEEEauthorblockN{Zahra Shakeri Hossein Abad\IEEEauthorrefmark{1}, Munib Rahman\IEEEauthorrefmark{1}, Abdullah Cheema\IEEEauthorrefmark{1}, Vincenzo Gervasi\IEEEauthorrefmark{2}, Didar Zowghi\IEEEauthorrefmark{3}, Ken Barker\IEEEauthorrefmark{1}}
    \IEEEauthorblockA{\IEEEauthorrefmark{1} Department of Computer Science, University of Calgary, Calgary, Canada
    \\\{zshakeri, murahman, cheemaab, kbarker\}@ucalgary.ca}
    \IEEEauthorblockA{\IEEEauthorrefmark{2}Department of Computer Science, University of Pisa, Italy, gervasi@di.unipi.it}
    \IEEEauthorblockA{\IEEEauthorrefmark{3}School of Software, University of Technology Sydney, Australia, didar.zowghi@uts.edu.au}
}

\makeatletter
\setcounter{figure}{-2}
\let\@oldmaketitle\@maketitle
\renewcommand{\@maketitle}{\@oldmaketitle
\includegraphics{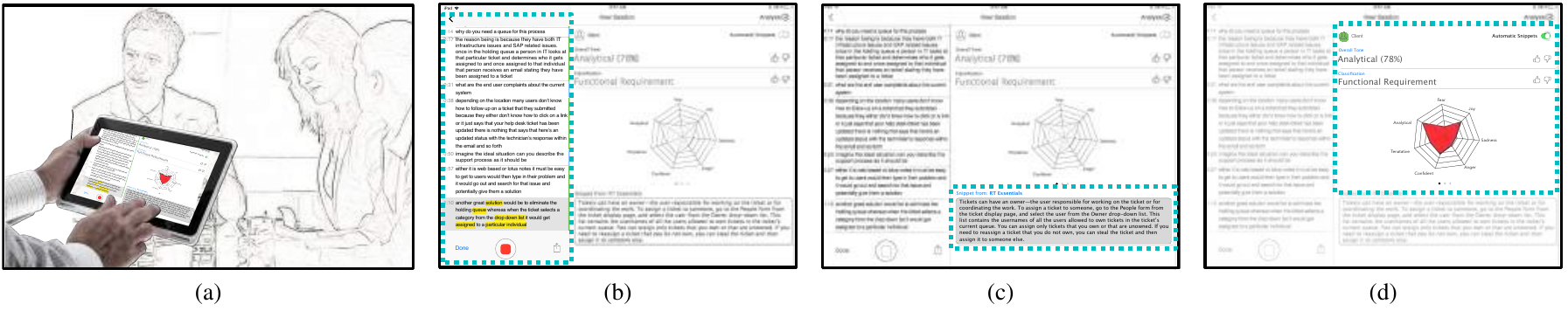}
    \vspace{-5mm}
\captionof{figure}{\bf (a) The application of ELICA during the collaboration with Stakeholders, (b) diarization feature included in the speech to text component, (c) extraction bar, which represents the extracted relevant information, (d) visualization and analytical component. }
  
  \label{fig:All}
  
   }
    
\makeatother
\maketitle

\begin{abstract}
Requirements elicitation can be very challenging in projects that require deep domain knowledge about the system at hand. As analysts have the full control over the elicitation process, their lack of knowledge about the system under study inhibits them from asking related questions and reduces the accuracy of requirements provided by stakeholders. We present {\em ELICA}, a generic interactive visual analytics tool to assist analysts during requirements elicitation process. ELICA uses a novel information extraction algorithm based on a combination of Weighted Finite State Transducers (WFSTs) (generative model) and SVMs (discriminative model). ELICA presents the extracted relevant information in an interactive GUI (including zooming, panning, and pinching) that allows analysts to explore which parts of the ongoing conversation (or specification document) match with the extracted information. In this demonstration, we show that ELICA is usable and effective in practice, and is able to extract the related information in real-time. We also demonstrate how carefully designed features in ELICA facilitate the interactive and dynamic process of information extraction. 

  \end{abstract}


\IEEEpeerreviewmaketitle

\section{Introduction and Motivation}
Requirements elicitation is an information-intensive task, drawing on different sources of information to infer a consensus on the requirements of a system. There have been numerous attempts to improve the flow of information, knowledge, or decision in elicitation process (e.g. \cite{REETA, Tool1}). However, two major obstacles still stand in the way of providing an effective communication between
 the systems analyst and clients: First, analysts and
 users often have different mental models and use different terminologies, filling the knowledge gap between the domain experts and the analysts during the requirements elicitation process is itself an open research problem.  Second, even if we had a perfect tool that could fully support analysts with requirements-related information, extracting and classifying the extracted information in real-time still remains challenging since this requires a dynamic modeling and analyzing the input text. For example, consider a situation in which an analyst is tasked to work on the requirements for a project from an unfamiliar domain. The misinformation during the requirements elicitation meeting reduces the accuracy of the requirements provided by stakeholders \cite{MisInfo}.

This paper describes ELICA (requirements ELICitation Aid), a tool for extracting and classifying requirements-relevant information during the elicitation process. Our focus is on both challenges: given an existing domain repository, how to extract requirements-relevant information in real-time. ELICA leverages the existing domain repositories as well as machine learning techniques to assist analysts to access relevant information about the application domain in real-time (Fig. \ref{fig:All}a). In ELICA, we use an off-the-shelf API---IBM Watson \cite{IBM}--- to convert spoken utterances to text strings and to diarize meeting data (Fig. \ref{fig:All}b). To extract requirements-relevant information (Fig. \ref{fig:All}c) we design and implement a hierarchical language model to measure the {\em lexical association} \cite{Lex2} between the latest window of a conversation (or text snippet) and the existing domain repositories. To extract this information in real-time, we apply Weighted Finite State Transducers (WFSTs) \cite{WFST}, powerful technique which handle variable-length feature vectors and support dynamic modeling and analysis of textual data. We leveraged the closure properties of WFSTs and the well-defined and flexible algebraic operations for these tools to enhance the SVMs technique for dynamic classification of the extracted requirements-relevant information (Fig. \ref{fig:All}d).

\section{ELICA Prototype}
\begin{figure}
\centering
\vspace{-5mm}
\includegraphics[scale=0.6]{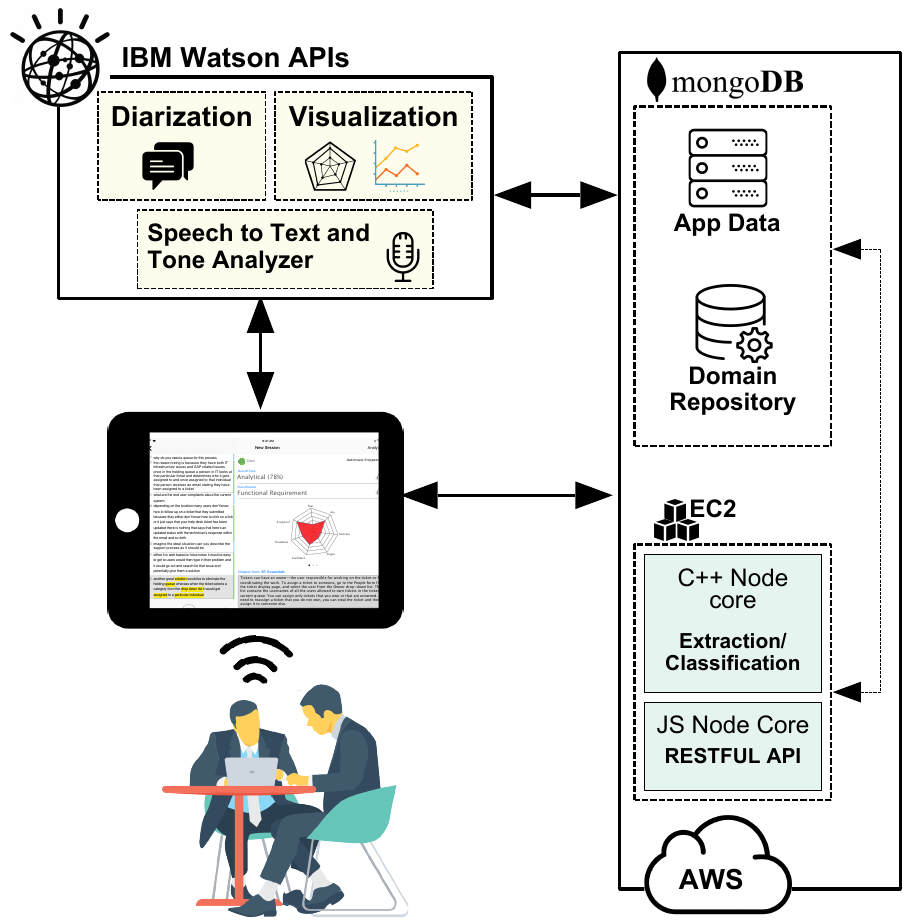}
\vspace{-4mm}
\caption{The overall architecture of ELICA.  }
\label{fig:Architecture}
\vspace{-5mm}
\end{figure}

The architecture of ELICA is shown in Fig. \ref{fig:Architecture}.  The entire ELICA system consists of two parts, the first part is responsible for extracting and classifying requirements relevant information, including {\em dynamic generative model} (using WFSTs) and {\em real-time classifier} (using SVMs); the other provides support for recording data and interaction with the extracted information, including {\em speech to text}, {\em speech diarization}, and {\em visualization} components. A typical user interaction process in ELICA engages both analytical and interactive components. 

{\bf Data Repository:} ELICA stores the domain repository and the application data (e.g. users' information, diarized conversation, and extracted information) in MongoDB on Amazon EC2, which communicates in JSON. ELICA is able to keep the data up-to-date by running the diarization, tone analyzer, extraction, and classification components in real-time.

{\bf Extracting and Classifying requirements:} Once the conversation between the client and the analyst has been initiated and transcribed, the most recent window of the conversation, along with the existing domain repository will be used as inputs to the information extraction module. After building the hierarchical language models, this module calculates the intersection between the two language models and returns the ``relevant'' terms which can be used to measure the lexical association between the on-going conversation and the existing repositories and fetch those parts of the domain repository that contain those relevant terms. The analyst can then interactively select among the returned text snippets those which best describe the on-going discussion.
ELICA not only returns the extracted relevant snippets but also classifies these snippets to Functional/Non-Functional Requirements (NFRs) as well as sub-categories of NFRs (e.g. usability, availability, security). 

{\bf Interactive Visualization:} ELICA offers an interactive interface that combines intuitive visualization about speakers' intentions such as confidence level, analytical tone, and emotions with a range of visual-textual information such as diarized speech (using a different color for each speaker) and extracted relevant information (highlighting content-carrying terms). Fig. \ref{fig:Architecture} outlines these components which are included in the {\em IBM Watson APIs} module. As the association between each recorded utterance and its relevant extracted snippet is stored in the App data storage, an analyst will be able to use this information at any time during/after the conversation and can easily access and review those parts of the conversation history (Fig. \ref{fig:All}b) that are of interest at the moment. Facilities that allow the user to rate and analyze the results of the data analysis are also provided (Fig. \ref{fig:All}d). 

{\bf Application Scenarios:}
The main three potential scenarios of ELICA are: {\bf [S1: Elicitation Meetings]:} An analyst is assigned to work on the requirements for a project from an unfamiliar domain. To support them during their elicitation meeting, ELICA selects inside each document of domain repository those textual snippets that are most relevant for the most recent part of the conversation; {\bf [S2: Resumption/Information Recall]:} An analyst is working on a requirements exploration task (e.g. elicitation meeting, discussion with other team members, ...) and needs to pause this task to address an incoming task. ELICA uses the recorded textual data during the exploration task as well as the existing problem domain documents to assist analysts better manage the issues of memory recall \cite{RE17, EASE18} after resuming the elicitation task; and {\bf [S3: Elicitation from Existing Documents]:} An analyst is assigned to an ongoing project and they need to become familiar with a possibly large amount of documentation that has already been produced. ELICA uses the existing problem domain documents to select textual snippets that are most relevant to the part of the system under investigation.

\section{Conclusion and Future Work}
The ELICA tool was developed to achieve two main goals. The first was to propose an effective tool for extracting requirements-relevant information during the elicitation process. This tool records, analyzes, and synthesizes the existing repository documents and incoming textual data in real-time and enables easy exploration of relevant information. Second, once ELICA has extracted and classified the relevant information, it supports analysts to interact with textual and visual results to increase their engagement with the extracted information and enhance their exploration abilities. So far, both components of the ELICA tool has been evaluated with an industrial dataset, yielding promising results, and we plan to further evaluate the tool in real development environments.

\printbibliography

\end{document}